\documentclass[prl,epsfig,a4paper,twocolumn,noshowpacs,floatfix,groupedaddress,
amsmath,amsfonts,amssymb,preprintnumbers,nofootinbib,citeautoscript]{revtex4}
\usepackage[T1]{fontenc}
\usepackage[latin1]{inputenc}
\usepackage{amssymb,amsmath,bm,dcolumn,graphicx}
\usepackage{color,booktabs,wrapfig,microtype,afterpage} \graphicspath{{Figures/}}
\usepackage[charter,greekuppercase=italicized]{mathdesign}
\usepackage{sidecap}
\usepackage[pdftex]{dropping}

\makeatletter\renewcommand{\fnum@figure}[1]{\textbf{\sffamily\figurename~\thefigure~|\,}}\makeatother
\makeatletter\renewcommand{\fnum@table}[1]{\textbf{\sffamily\tablename~\thetable~|\,}}\makeatother

\newcommand{\Rr}{\frac{R}{r}}

\newcommand{\ellR}{\frac{\ell}{R}}
\newcommand{\ellmax}{\ell_\text{max}}

\newcommand{\ellavgR}{\frac{\langle \ell \rangle}{R}}

\definecolor{NatureBlue}{rgb}{0.012,0.3,0.63}
\definecolor{NiceOrange}{rgb}{0.85,0.42,0.21}
\usepackage[colorlinks,plainpages=false,linkcolor=NatureBlue,urlcolor=NatureBlue,
citecolor=NatureBlue,pdfpagemode=UseNone,pdfstartview=FitBH]{hyperref}

\makeatletter
\newcommand{\bibstyle@supplement}{\bibpunct[, ]{[S}{]}{;}{n}{,}{,}%
    \gdef\bibnumfmt##1{[S##1]}}
\makeatother

\begin{document}

\title{\flushleft\fontsize{15.8pt}{15.8pt}\selectfont\sffamily \textcolor{NiceOrange}
{Compaction of Quasi One-Dimensional Elastoplastic Materials}}

\author{\sffamily M. Reza Shaebani$^1$, Javad Najafi$^{2,3}$, 
Ali Farnudi$^3$, Daniel Bonn$^{4}$ \& Mehdi Habibi$^{3,4,*}$\smallskip}

\affiliation{\flushleft
\mbox{\sffamily $^1$\hspace{0.5pt}Department of Theoretical Physics,
Saarland University, 66041 Saarbr\"ucken, Germany}
\mbox{\sffamily $^2$\hspace{0.5pt}Department of Experimental Physics,
Saarland University, 66041 Saarbr\"ucken, Germany}
\mbox{\sffamily $^3$\hspace{0.5pt}Department of Physics, Institute
for Advanced Studies in Basic Sciences, Zanjan 45195, Iran}
\mbox{\sffamily $^4$\hspace{0.5pt}Van der Waals-Zeeman Institute,
University of Amsterdam, 1098 XH Amsterdam, The Netherlands}
\mbox{\sffamily $^*$\hspace{0.5pt}Current address: Laboratory of Physics and Physical 
Chemistry of Foods, Wageningen University, Wageningen, The Netherlands}
\mbox{Correspondence and requests for materials should be addressed to 
D.\,B. (email: D.Bonn@uva.nl), M.\,H. (email: }
\mbox{M.Habibi@uva.nl), or M.\,R.\,S. (email: shaebani@lusi.uni-sb.de).}}

\begin{abstract}\citestyle{nature}
\parfillskip=0pt\relax\fontsize{9pt}{11pt}\selectfont\noindent\textbf{
Insight in the crumpling or compaction of one-dimensional objects 
is of great importance for understanding biopolymer packaging and 
designing innovative technological devices. By compacting various 
types of wires in rigid confinements and characterizing the morphology 
of the resulting crumpled structures, here we report how friction, 
plasticity, and torsion enhance disorder, leading to a transition 
from coiled to folded morphologies. In the latter case, where folding 
dominates the crumpling process, we find that reducing the relative 
wire thickness counter-intuitively causes the maximum packing density 
to decrease. The segment-size distribution gradually becomes more 
asymmetric during compaction, reflecting an increase of spatial 
correlations. We introduce a self-avoiding random walk model and 
verify that the cumulative injected wire length follows a universal 
dependence on segment size, allowing for the prediction of the efficiency 
of compaction as a function of material properties, container size, 
and injection force.}
\end{abstract}

\pagestyle{plain}
\makeatletter
\renewcommand{\@oddfoot}{\hfill\bf\scriptsize\textsf{\thepage}}
\renewcommand{\@evenfoot}{\bf\scriptsize\textsf{\thepage}\hfill}
\makeatother

\citestyle{nature}
\maketitle

\makeatletter\immediate\write\@auxout{\string\bibstyle{my-nature}}\makeatother
\renewcommand\bibsection{\section*{\sffamily\bfseries\footnotesize References\vspace{-10pt}\hfill~}}

\dropping[0pt]{2}{C}ompaction of slender objects in confined
geometries is ubiquitous in nature. Perhaps the most important
example is DNA packaging in viral and bacteriophage capsids 
and cell nuclei \cite{Purohit03,Ali06,Katzav06,Kindt01,Smith01}. 
Other pertinent examples are the folding of insect wings 
in cocoons \cite{Brackenbury94}, and flower or plant leaves in 
buds \cite{Kobayashi98}. The process of compaction may result 
in complex morphologies depending on the applied forces and 
constraints \cite{Deboeuf13}. Recent numerical studies 
\cite{Vliegenthart06,Tallinen09,Balankin09,Balankin13,Sultan06} 
showed that the space-filling properties of 2D crumpled 
sheets are influenced by parameters such as self-avoidance 
and plasticity - ingredients that are difficult to disentangle 
in experiments. Self-avoidance alters the hierarchical nature 
of the compaction process and induces stronger self-correlations 
as the compression increases \cite{Sultan06,Bayart14}. Thus, 
considering the structural evolution is key for understanding 
the efficiency of \emph{in vitro} compaction. 

Aiming to provide quantitative insights into the role of self-avoidance, 
we turn to 1D wires. Due to its very nonlinear nature it is easier 
to study 1D systems than the more complicated crumpling process 
in 2D sheets \cite{Sultan06,Lin09,Cambou11,Aharoni10}. For 
1D-compaction, how the morphology of crumpled objects develops 
is of particular importance in technological and biological 
applications as, for example, in endovascular coiling treatment 
of cerebral aneurysms \cite{aneurysm} or in packing of DNA \cite{Witten98}. 

\begin{figure*}
\includegraphics[width=0.99\columnwidth]{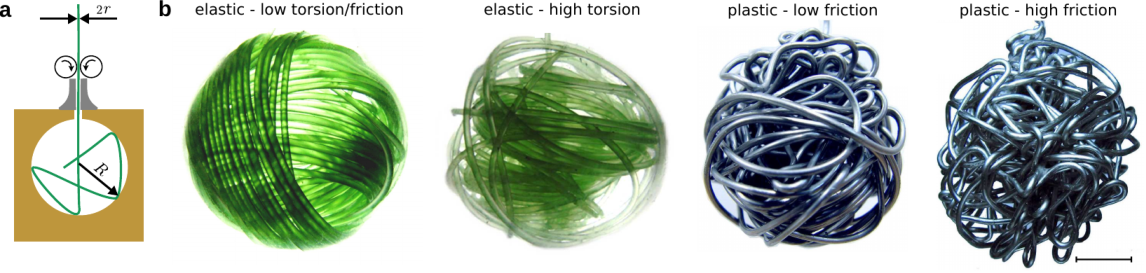}
\caption{\textbf{Experimental setup and examples of distinct morphologies.} 
(\textbf{a}) Cross-sectional view of the experimental setup used to compact 
1D wires in a spherical rigid container. (\textbf{b}) Distinct morphologies 
obtained by compacting wires with different material properties introduced 
in the \emph{Methods} section. Scale bar, $1 \text{cm}$.
\label{Fig1}}
\end{figure*}

When compacting elastic low-frictional wires with a high bending 
rigidity in confined geometries in such a way that the internal 
torsion is released, highly ordered structures with distinctly 
oriented subdomains of parallel coils form (Fig.\,\ref{Fig1}). 
However, with increasing friction \cite{Vetter14} or plasticity 
\cite{Tallinen09}, or by accumulating torsion during the packing 
process \cite{Stoop11,Najafi12}, disordered structures emerge 
where the contribution of folds or bends in the morphology is 
more pronounced. For example, by introducing the number of 
segments as the order parameter, it has been recently shown 
that a sharp transition from ordered (coiled) to disordered 
(folded) structures occurs as the friction increases \cite{Vetter14}. 
In disordered morphologies, the compaction efficiency is controlled 
to a large extent by spatial exclusion effects, which continuously 
evolve in the course of compaction. Hence, unraveling the    
\onecolumngrid

\clearpage

\twocolumngrid
\onecolumngrid

\begin{figure*}
\centering
\includegraphics[width=0.99\columnwidth]{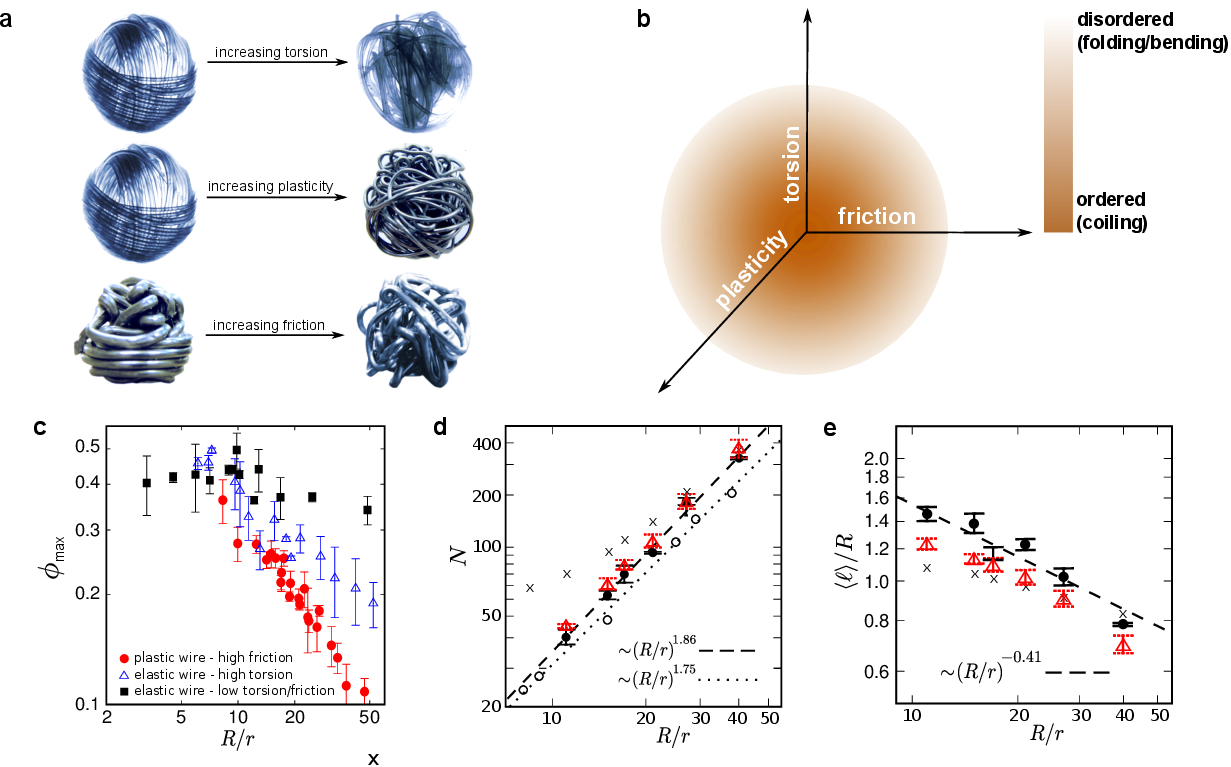}
\caption{\textbf{Morphological phase diagram and scaling of 
geometrical characteristics.} (\textbf{a}) Examples of morphological 
differences upon increasing key material parameters. (\textbf{b}) 
Schematic morphological phase diagram in the space spanned by torsion, 
friction, and plasticity. (\textbf{c}) The maximum packing density 
$\phi_\text{max}$ vs the relative system size $\Rr$ (parameters 
defined in Fig.\,\ref{Fig1}a). (\textbf{d,e}) Scaling of the number 
of bends $N$ (d), and the dimensionless mean segment size $\frac{
\langle\ell\rangle}{R}$ (e) versus the effective system size $\Rr$, 
for plastic frictional wires. The filled circles (open triangles) 
indicate experimental (simulation) results. The dashed lines 
are power-law fits to the experimental data. The crosses denote 
the simulation results obtained when assuming that the packing 
fraction is the influential parameter on the self-avoidance 
effects. The open circles in panel (d) show the experimental 
results obtained from the high frictional setup. Error bars 
correspond to standard deviation of $5$ separate measurements.}
\label{Fig2}
\end{figure*}

\twocolumngrid
\noindent 
mechanisms that govern the evolution of self-avoidance is crucial to 
achieve an efficient compaction. 

Here we study the morphologies of wires packed into rigid spherical 
containers and interestingly find that the maximum packing density 
in disordered structures decreases with reducing the thickness 
of the wire (or, equivalently, increasing the container size). 
To elaborate on the underlying mechanisms leading to this peculiar 
behavior, we isolate the influence of self-avoidance by focusing 
on the compaction of plastic frictional wires where folding is 
dominant in the resulting structure. By following the morphological 
evolution, a gradual crossover from random to correlated folding 
events is observed due to spatial exclusion effects. We propose 
that the compaction can be considered as a confined self-avoiding 
random walk (SAW). In such far-from-equilibrium processes, the 
imposed constraints and initial conditions do not uniquely determine 
the final crumpled state. Instead, there is an ensemble of admissible 
configurations, from which some structural properties of the system 
can be derived. We introduce a SAW sampling method which successfully 
accounts for the time evolution of the wire segment length. We thus 
present a complete understanding of the compaction: the maximum 
length of the injected wire can be estimated from the geometry and 
imposed constraints for a given set of material parameters.

\smallskip
\smallskip
\noindent\textbf{Results}

\noindent\textbf{Universal phase diagram for 1D crumpling.} 
We first consider the packing of elastic low-frictional wires with a high 
bending rigidity in rigid spherical containers. When the wire is allowed 
to axially rotate at the injection point to release the torsion during 
the packing process, highly ordered coils form as the wire relaxes 
towards a global minimum energy. By hindering the release of 
torsion, the wire buckles more frequently to free elastic energy. 
Hence, the packing process becomes less ordered \cite{Stoop11}, 
leading to warped structures similar to those obtained numerically 
for compaction of DNA molecules in phage capsids \cite{Kindt01} 
(see Figs.\,\ref{Fig1} and \ref{Fig2}a). The disorder is also 
enhanced by friction, which causes the wire to resist against 
sliding and to randomly bend due to local constraints \cite{Vetter14}. 
Another property which obviously affects the morphology, is 
the degree of plasticity of wire. While the bending rigidity 
of the plastic wires can be quite high, their yield stress is 
relatively low, leading to structures with rather straight segments 
and sharp turnings. Upon increasing plasticity (i.e.\ lowering 
the yield stress), the irreversible deformations of wire increase 
the disorder of the crumpled configuration. One can map out a qualitative 
phase diagram for the morphological evolution of the resulting 
crumpled structures in the space of wire properties (friction, 
torsion, plasticity), as depicted in Fig.\,\ref{Fig2}b. More 
generally, disordered structures can be generated in diverse 
ways by tuning the wire or container properties. The 
morphological phase space indeed contains additional degrees 
of freedom associated with container properties, such as 
its flexibility \cite{Vetter14}, shape \cite{Petrov07}, or 
the degree of confinement imposed by it (characterized by 
the container size $R$ relative to the radius of gyration 
$R_g$ of the crumpled structure and also to the persistence 
length $\lambda$ of the elastic wire). For example, a biopolymer 
coils itself inside the cage if $\lambda$ is comparable 
to $R$ (e.g.\ in packing of DNA in icosahedral bacteriophages), 
while for weak confinement, i.e.\ $\lambda{\ll}R_\text{g}{<}R$, 
the biopolymer chain (such as chromatin) has a relatively 
low bending stiffness and behaves as a self-avoiding random 
walker without `feeling' the boundaries. For $\lambda$ values 
in between, the morphological evolution during the crumpling 
process is complicated due to varying combined effects of 
self-avoidance and interactions with boundaries \cite{Witten98,Odijk08}. 
It has been also shown that a transition from coiled to 
disordered configurations occur, as the accessible space 
reduces during the compaction of elastic rods \cite{Pineirua13}.

To compare the compaction efficiencies, we measure the packing 
density once the injection of wire eventually stops. Indeed, 
the value obtained for this maximum packing density, 
$\phi_\text{max}$, depends on the wire radius $r$, the 
container size $R$, and the insertion force. We measure 
this quantity for a given insertion force and for different 
combinations of inserted wire radius $r$ and container radius 
$R$. When plotting $\phi_\text{max}$ vs the non-dimensional 
system size $\Rr$ for the coiled compact morphologies of 
low-plasticity, low-friction, low-torsion wires 
(Fig.\,\ref{Fig2}c), a plateau at small $\Rr$ followed by 
a weak decrease at larger values of $\Rr$ is observed. It 
was shown with geometrical arguments \cite{Najafi12} that 
$\phi_\text{max}$ slightly decays with $\Rr$ for a purely 
coiled structure in a spherical container. Note that the 
very inner core of the structure practically becomes 
disordered as the accessible space reduces and its shape 
becomes more irregular (which makes the formation of coils 
more difficult). This disordered core (with a possible 
$\Rr$-dependent size) can also contribute to the weak decay 
of $\phi_\text{max}$ vs $\Rr$. It has been shown that the 
packing fraction decreases with increasing disorder in 
packings of elastic wire \cite{Bayart11}. 

When increasing plasticity, friction and/or torsion, resulting 
in the formation of folds and bends, the data collapse onto 
curves following a power-law $\phi_\text{max}{\sim}(R{/}r)^{-
\alpha{=}D{-}3}$, with $D$ being the fractal dimension \cite{Donato02,
Donato03}. The slope of the curve depends on the degree of disorder. 
For example, lubricating the inner wall of the container with 
silicon oil leads to the formation of highly ordered coils 
at the outer layer of crumpled plastic wires, which results 
in a mixed coiled-folded structure with $\alpha{\simeq}0.38
{\pm}0.03$. A similar exponent is obtained for the compaction 
of elastic torsional wires where coils and bends coexist. 
The steepest descent is observed for crumpling of plastic 
wires at high friction where coils are absent and folding 
is the dominant process ($\alpha{\simeq}0.52{\pm}0.05$). 
While the very weak system-size dependence of efficient 
compaction in ordered (coiled) structures is understandable, 
the behavior of disordered morphologies is counter-intuitive, 
as one would expect that relatively thinner wires more flexibly 
fill a given container, leading to a higher compaction 
efficiency. A similar trend for the dependence of packing 
density on the relative system size was reported in 
experiments on DNA packaging in viral capsids \cite{Purohit05}, 
revealing that in spite of the huge differences in length 
scales of the two systems, the maximum packing densities 
behave similarly in the presence of disorder. Self-avoidance 
inside a confinement can explain the peculiar behavior of 
$\phi_\text{max}$ versus $\Rr$ via a mean-field interpretation, 
assuming that the self-avoidance energy originates mainly 
from the homogeneously distributed binary contacts between 
the wires (whose density nearly grows as the square of the 
packing fraction), and also supposing that the local radius 
of curvature of the confinement is comparable to the 
container size and varies slowly. When balancing the 
confining energy and self-avoidance \cite{deGennes85,
Kantor86,Gomes08}, the lowest (harmonic) approximation 
of the confining energy yields an energy density of the 
order of $(\Rr)^{-1}$ while the self-avoidance energy 
density is proportional to $\phi^2{\sim}(\Rr)^{-2\alpha}$. 
By equalizing these energy densities we obtain $\phi {\sim}
(\frac{R}{r})^{-0.5}$.

\smallskip\smallskip
\noindent\textbf{Segment-size statistics.} 
To better understand the influence of disorder, we choose plastic 
frictional wires to avoid ordered coils and create the 
highest possible disorder in the crumpled structure. After 
compacting the wires, we open the molds and investigate the 
resulting compacted structures by analyzing the folding 
statistics. The points of folding were often determined by 
sharp changes of wire orientation. If they were not easily 
distinguishable, then a minimum threshold of $90$ degrees 
for the turning angle of the wire, and a maximum threshold 
of $\frac{R}{2}$ for the radius of curvature were imposed. 
We cut the wire at each of the folding points, straighten 
the segments, and measure their length. Straightening of 
the curved segments rarely allows for segment lengths $\ell$ 
longer than the container diameter, but we checked that 
the maximum segment length $\ellmax$ remains smaller than 
$\pi R$ in the absence of coils. We preserve the order 
of the wire segments and average the results over 5 
realizations for each value of $R{/}r$ to obtain the 
sequence of the segment lengths $\ell_n$.

A key observation is the scaling of the total number of segments
$N$ with the effective system size $\Rr$. Similar scaling laws 
were reported for 2D packings of wires \cite{Donato02,Gomes08,
Stoop08}. As shown in Fig.\,\ref{Fig2}d, a power-law relation 
of the form
\begin{eqnarray}
N\,{\sim}\,(R{/}r)^\beta
\label{Eq:N-scaling}
\end{eqnarray}
holds with $\beta{=}1.86{\pm}0.03$ for smooth wire and container 
with wire-wire and container-wire friction coefficients $\mu_\text{ww}{
\simeq}0.2$ and $\mu_\text{cw}{\simeq}0.4$, respectively. The 
exponent can be understood by considering the wire crumpling 
process as a self-avoiding random walk (SAW) in confinement. 
For comparison, for the number of steps on a cubic lattice in 
an ordinary random walk $\beta$ equals 2, while for SAW 
$\beta{\simeq}5{/}3$ \cite{Havlin82}. The fact that we find 
an exponent in between these two values can be understood 
because there is a gradual evolution of spatial correlations 
over the course of crumpling (see below), thus, the exponent 
continuously decreases from $2$. However, the wires can slide 
over each other due to the finite friction so that the 
self-avoidance constraint is only partially fulfilled. For 
comparison, we repeated the experiment by roughening the 
plastic wires and the inner surface of the molds to increase 
the friction coefficients to $\mu_\text{ww}{\simeq}0.45$ 
and $\mu_\text{cw}{\simeq}0.45$. The considerable change 
in the wire-wire friction resulted in a smaller exponent 
$\beta{=}1.75{\pm}0.05$ which is closer to the pure self-avoidance 
limit (see Fig.\,\ref{Fig2}d). From the scaling of $\phi_\text{max}$ 
and $N$ with $\Rr$ one expects that the normalized mean 
segment size $\ellavgR$ follows $(\Rr)^{-0.4}$, as 
confirmed by the experimental results in Fig.\,\ref{Fig2}e. 

\begin{figure}[b]
\centering
\includegraphics[width=0.99\columnwidth]{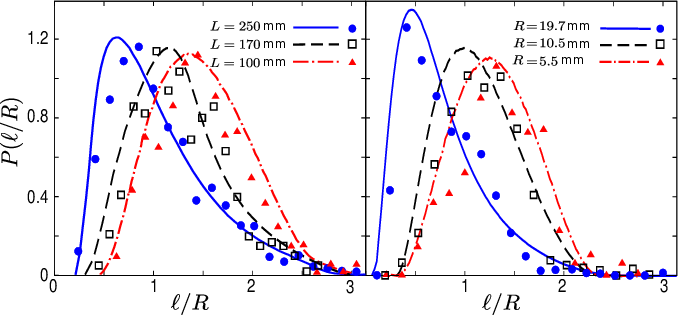}
\caption{\textbf{Probability distribution of the normalized 
segment size $\ellR$.} The symbols (lines) represent the 
experimental (simulation) results for $r{=}0.5\,\text{mm}$. The 
results in the left panel are obtained by injecting wires of 
different total length $L$ into a container with radius $R{=}14.5\,
\text{mm}$. The right panel represents the results of injecting 
the maximum possible length of wire (when applying the insertion 
force of nearly $F{=}100\,N$) in different container sizes $R$.}
\label{Fig3}
\end{figure}

\smallskip\smallskip
\noindent\textbf{Evolution of spatial correlations.} 
At earlier stages of the crumpling process in a given mold, the 
injected wire proceeds in the container without interacting with 
the accumulated wire. Assuming that the plastic wire bends at 
a random point between the injecting hole and a contact point 
at the container surface, the resulting segment length $\ell$ 
is a random variable, symmetrically distributed between $0$ 
and the maximum possible segment length $\ellmax$. By increasing 
the total length $L$ of accumulated wire, spatial exclusion 
effects grow and the injected wire cannot easily proceed through 
the sphere without touching the crumpled structure. Hence, long 
segments gradually become less probable and the probability 
distribution $P(\ellR)$ of the normalized segment size becomes 
more asymmetric due to relatively large populations of smaller 
segments (see Fig.\,\ref{Fig3}a). When comparing the final 
structures (i.e.\ those obtained when the injection of wire 
stops), we interestingly find that for larger values of $\Rr$ 
the segment size distribution $P(\ellR)$ is more asymmetric 
and shifts towards smaller segment sizes (Fig.\,\ref{Fig3}b). 
This behavior similarly indicates the growth of spatial exclusion 
effects with increasing $\Rr$. Note that the initial segment 
sizes $\ell$ are only determined by the container size $R$ in 
all containers, however, $\ell$ gradually decreases as the 
spatial exclusion effects grow. The effect is more pronounced 
for larger spheres as the crumpling process continues further.

\begin{figure}[b]
\centering
\includegraphics[width=0.99\columnwidth]{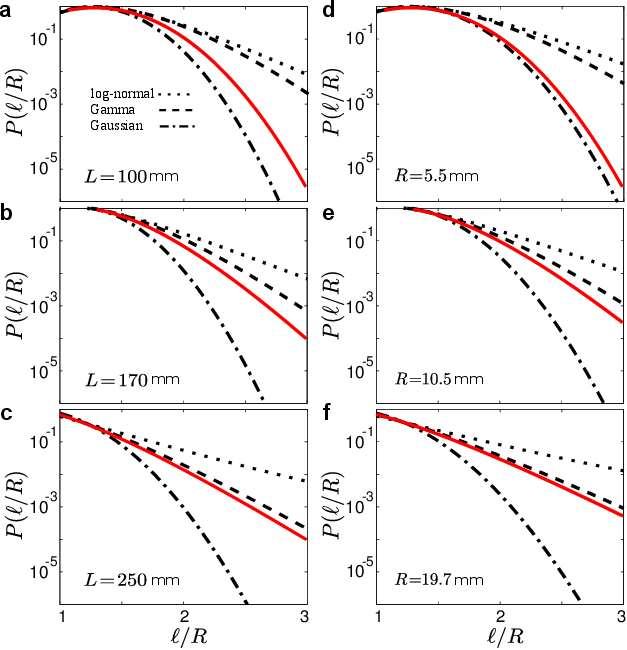}
\caption{\textbf{Evolution of the tail of 
the probability distribution $P(\ellR)$.} The results obtained 
from the simulations for $r{=}0.5\,\text{mm}$. (\textbf{a-c}) Comparison 
between the tail of $P(\ellR)$ for different values 
of the total length $L$ of the injected wire in a container 
with radius $R{=}14.5\,\text{mm}$. (\textbf{d-f}) A similar comparison 
when the maximum possible length of wire is injected in 
different container sizes $R$. The solid lines denote the 
simulation results, and the dotted, dashed, and dashed-dotted 
guidelines represent, respectively, the log-normal $P(x)
{=}\frac{1}{\sqrt{2\pi}\sigma{x}}\exp[-\frac{(\ln x{-}
\lambda)^2}{2\sigma^2}]$, gamma $\frac{x^{\sigma{-}1}}{
\lambda^\sigma\Gamma(\sigma)}\exp[-x{/}\lambda]$, and 
Gaussian $\frac{1}{\sqrt{2\pi}\sigma}\exp[-\frac{(x{-}
\lambda)^2}{2\sigma^2}]$ distributions, plotted with the 
same mean and variance as the corresponding experimental 
data. In the gamma distribution, $\Gamma(\sigma)$ denotes 
the gamma function evaluated at $\sigma$.}
\label{Fig4}
\end{figure}

\smallskip\smallskip
\noindent\textbf{Self-avoiding random walk (SAW) model.} 
We argue that the strength of self-avoidance effects is indeed 
captured by the total length $L$ of the injected wire, rather 
than the total volume excluded by it (i.e.\ the packing fraction). 
The exclusion effect that an inserted rod-like object experiences 
inside a crumpled structure is effectively determined by the projection 
of the crumpled wire on a plane perpendicular to the direction of 
insertion. Therefore, the total length and the thickness of the 
crumpled wire are expected to be the influential parameters. 
However, the circular cross section of wires reduces the contact 
area between the touching wires and, thus, the effective frictional 
force between them. As a result, the self-avoidance effects are 
not proportionally increasing with the wire thickness (i.e.\ $r$). 
We conclude that the entire contribution to the spatial exclusion 
constraint can be attributed to the length of wire, reflected in 
the dimensionless quantity $\lambda{=}\frac{L}{R}$ which grows as 
$\lambda{\sim}(\Rr)^{1.5}$ (while $\phi$ decreases as $\phi{\sim}
(\Rr)^{{-}0.5}$). In the following, we simulate the folding 
process as a SAW of the wire inside the confinement. While the 
existing SAW algorithms mainly follow stochastic Markovian 
dynamics to sample the ensemble of trajectories on regular 
lattices, here we propose an alternative approach which 
accounts for the time evolution of the step size $\ell$. We 
suppose that the strength of self-avoidance effects after 
the $n$-th segment is mainly controlled by the length $L_n$ 
of the inserted wire, i.e.\ the larger is the parameter 
$\lambda_n{=}\frac{L_n}{R}$, the smaller is the success 
probability for the segment $n{+}1$ to be a long one. The 
size $\ell_{n{+}1}$ of the next segment is obtained via 
the following algorithm: A trial segment size $\ell$ is 
chosen randomly within $[0,\ellmax]$, with $\ellmax$ 
being the maximum segment length obtained in experiments 
for a given value of $R$. The proposed $\ell$ is accepted 
according to a Metropolis-like criterion with probability
\begin{eqnarray}
P\!\!_{n{+}1}(\ell)=\mathcal{N}^{-1}\exp\big[{-}\kappa\lambda_n\,\ell{/}R \big],
\label{Eq:AcceptanceProb}
\end{eqnarray}
where $\mathcal{N}{=}\frac{R}{\kappa \lambda_n} (1{-}\exp[-\kappa
\lambda_n \ell_\text{max}/R])$ is the normalization factor.
The coefficient $\kappa$ depends on wire properties and is
treated as a free parameter to take into account the partial
fulfillment of the self-avoidance constraint due to sliding
of the wires. While the exponents are not affected by the
choice of $\kappa$, by fitting it we can quantitatively
reproduce the experimental data. For the sake of simplicity, 
here we used a single averaged value of $\kappa$ to reproduce 
all the experimental data. In the case of rejection, a new
$\ell$ is tried. Finally, the cumulative length is updated 
as $L_{n{+}1}{=}L_n{+}\ell$ before starting the next step. 
Equation(\ref{Eq:AcceptanceProb}) assumes an exponentially 
lower acceptance chance for larger trial lengths $\ell$. 
Moreover, the acceptance probability decreases with increasing 
$\lambda_n$, as the self-avoidance effects become more pronounced. 
The method samples the segment-length landscape according 
to a Boltzmann-like distribution. Initially, the wire walks 
in free space ($\lambda_n{=}0$), thus, $P\!\!_{n{+}1}(\ell)
{=}\frac{1}{\ell_\text{max}}$ independently of the trial 
length $\ell$. However, larger values of $\ell$ become 
gradually less probable with increasing $\lambda_n$. We 
perform extensive Monte Carlo simulations by adjusting 
$\ellmax$ and the threshold value of $\phi$ to the experimental 
data. The results shown in Figs.\,\ref{Fig2},\,\ref{Fig3} 
are in remarkable agreement with experiments. We checked 
that the power-law scalings can not be reproduced when 
replacing $\lambda_n$ with $\phi_n$ in Eq.\,(\ref{Eq:AcceptanceProb}) 
(see Figs.\,\ref{Fig2}d and \ref{Fig2}e). Notably, the 
numerical predictions for the total wire length $L$ differ 
less than $4\%$ from the experimental values for all system 
sizes. 

\begin{figure}[b]
\centering
\includegraphics[width=0.99\columnwidth]{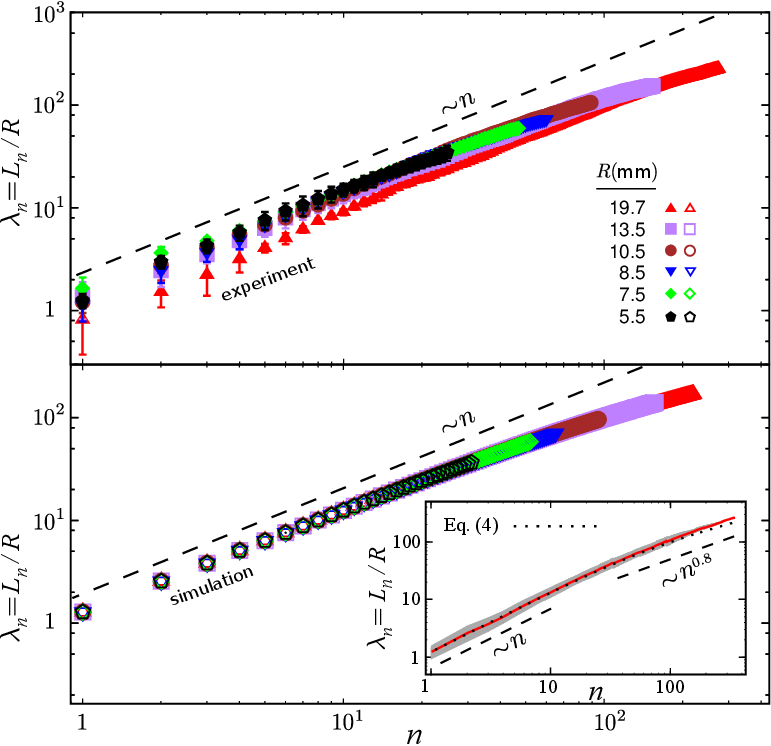}
\caption{\textbf{The cumulative length $L_n$ of the inserted 
wire after the $n$-th segment.} The top (bottom) panel represents 
experimental (simulation) results. (inset) $L_n{/}R$ averaged 
over all experiments (red solid line) and its standard deviation 
(gray shaded area). The dotted line is obtained from 
Eq.\,(\ref{Eq:Sum}).}
\label{Fig5}
\end{figure}

In Fig.\,\ref{Fig4} we take a closer look at the tail of 
$P(\ellR)$ obtained from the numerical simulations. While 
Gaussian function represents the distribution of random 
uncorrelated data, gamma and log-normal functions are 
respectively associated with random events in the presence 
of self-correlations and those that occur hierarchically 
\cite{Vliegenthart06,Sultan06,Wood02,Blair05,Adda-Bedia10,
Deboeuf13}. Since the noisy data of the experimental tail 
prevents any conclusive statement on the tail behavior 
of $P(\ellR)$, we plot these three functions (with the 
same mean and variance as the experimental data) and use 
them as guidelines to demonstrate the trend of the tail 
behavior obtained from the numerical simulations. We 
investigate the evolution of the tail during the injection 
of wire in a given container, and also compare the tails 
when the maximum possible length of wire is injected in 
different container sizes. As shown in Fig.\,\ref{Fig4}, 
the tail is better captured by the Gaussian for small 
$\Rr$ or at the early stages of crumpling, while there 
is a gradual crossover towards the Gamma distribution, 
either by increasing the container size or by increasing 
the length of the injected wire. Thus, self-correlations 
are the dominant underlying mechanism here. A hierarchical 
folding mechanism is expected to cause a rather stable 
log-normal distribution tail over all timescales, thus, 
the evolution of the tail is in favor of evolving correlated 
events. It has been previously shown numerically \cite{Sultan06} 
and by compacting of rods in 2D experiments \cite{Bayart14}, 
that self-avoidance alters the hierarchical nature of 
crumpling at high compression and induces self-correlations.

\begin{figure}[b]
\centering
\includegraphics[width=\columnwidth]{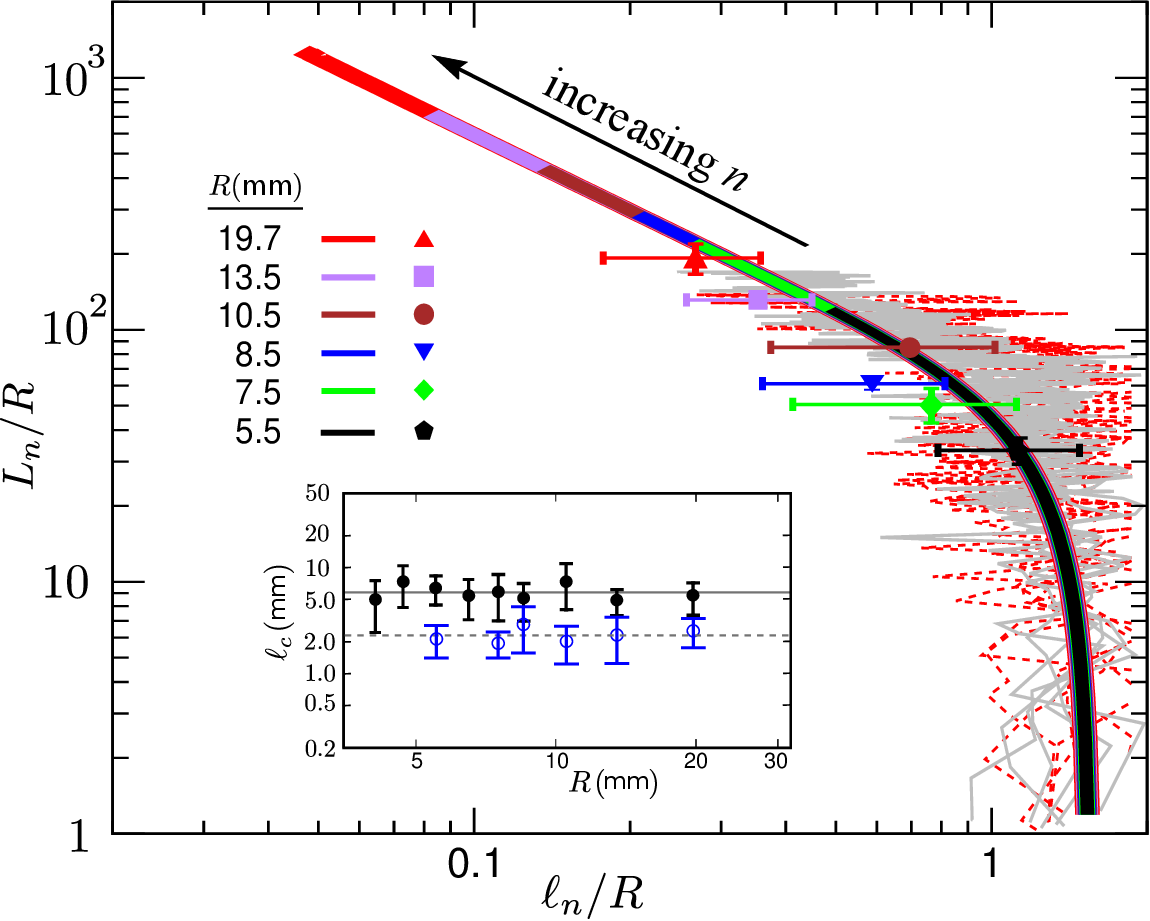}
\caption{\textbf{Universal filling mechanism.} Collapse of $L_n{/}R$ 
versus the scaled segment length $\ell_n{/}R$ is shown for different 
values of $\Rr$ in experiments (background gray lines) and simulations 
(thick colored lines). The symbols show the experimental cutoff 
values $\ell_c{/}R$ for $r{=}0.5\,\text{mm}$. The red dashed lines 
indicate the experimental results when injecting wires of different 
total length $L{\approx}100\,\text{cm}$, $170\,\text{cm}$, or $250\,
\text{cm}$ into a container with radius $R{=}14.5\,\text{mm}$. 
(inset) The cutoff segment length $\ell_c$ vs the container 
radius $R$ for $r{=}0.3\,\text{mm}$ (open circles) and $r{=}
0.5\,\text{mm}$ (filled circles). The horizontal lines show 
the average values $\bar\ell_c{=}2.3$ and $5.8\,\text{mm}$. 
Error bars correspond to standard deviation of $5$ separate 
measurements.}
\label{Fig6}
\end{figure}

The cumulative length $L_n$ of the inserted wire after the
$n$-th segment qualitatively collapses onto a master curve 
for different values of $\Rr$ (Fig.\,\ref{Fig5}). The segments 
are initially independent of each other and $L_n{/}R$ grows
linearly with $n$. The steps however become more correlated
with increasing $n$, leading to a slower growth of $L_n{/}R$.
A similar reduction of the slope has been recently observed 
for motor-driven viral packaging \cite{Keller16}. From the 
scaling of $\phi$ and $N$ with $\Rr$, one obtains the 
asymptotic scaling $L{/}R\,{\sim}\,N^{0.8}$, which is 
consistent with experiments [see Fig.\,\ref{Fig5}(inset)]. 
Starting from $\langle{\ell_1}\rangle{=}\frac{\ell_\text{max}}{2}$,
the mean segment size at next steps can be estimated in 
terms of $L_n$ as
\begin{eqnarray}
\langle{\ell_{n{+}1}}\rangle=\int_0^{\ell_\text{max}}
\!\!\!\!\!\!\!\!\!\!\!\! P\!\!_{n{+}1}(\ell)\,\ell\,
\text{d}\ell=\frac{R}{\kappa\lambda_n}{+}\frac{\ell_\text{max}}
{1{-}e^{\kappa\lambda_n\ell_\text{max}/R}}.
\label{Eq:ln-Average}
\end{eqnarray}
Hence, we obtain the following recursive analytical
expression for the injected length of wire after the
$n$-th segment
\begin{eqnarray}
\frac{L_n}{R}{=}\frac{\ell_\text{max}}{2R}{+}
\sum_{i{=}1}^{n{-}1}\bigg[\frac{R}{\kappa L_i}{+}\frac{\ell_\text{max}
{/}R}{1{-}\exp[\kappa\ell_\text{max}L_i{/}R^2]}\bigg],\,\,\,\,\,
\label{Eq:Sum}
\end{eqnarray}
in excellent agreement with the data as shown in 
Fig.\,\ref{Fig5}(inset).

\smallskip\smallskip
\noindent\textbf{Buckling threshold.} The inset of Fig.\,\ref{Fig6} 
shows that the injection of wire eventually stops at a cutoff 
segment length $\ell_c$ which is independent of $R$ for a given 
$r$. It follows that the process stops when the segment size 
becomes so small that the applied feeding force $F$ exceeds 
the maximum power of the motor. The minimum threshold length 
can be obtained from the Euler buckling theory as $\ell_c{=}
\sqrt{\frac{\pi^3Y}{16F}}r^2$ so that $\ell_c$ is determined 
by the insertion force and wire properties (Young's modulus 
$Y$) and is independent of $R$. For $r{=}0.3\,\text{mm}$ 
($0.5\,\text{mm}$) and a maximum insertion force of nearly 
$F{=}100\,N$ in our setup, we obtain $\ell_c{\simeq}2.2\,
\text{mm}$ ($6.0\,\text{mm}$), close to the experimental 
values of 2.3 and 5.8 mm [Fig.\ref{Fig6}(inset)].

Figure\,\ref{Fig6} interestingly evidences a universal filling
mechanism independent of $R$. The normalized cumulative wire
length $L_n{/}R$ versus the normalized length of the $n$-th
segment $\ell_n{/}R$ collapses onto a universal curve for
different values of $\Rr$. As the wire injection continues,
$\ell_n {/}R$ gradually decreases until it reaches the minimum
threshold value $\ell_c{/}R$, where the process eventually
stops. For a given value of $r$, $\ell_c$ is independent of
$R$, thus, $\ell_c{/}R$ decreases (i.e. shifts to the left
in Fig.\,\ref{Fig6}) with increasing $R$. Consequently, the
cutoff number of the steps at which the process stops
increases with $R$. By calculating $\ell_c$ from the insertion
force and wire properties, one can predict the total length
of crumpled wire for different $R$. The data collapse is also 
obtained when injecting wires of different total length $L$ 
into a given container (dashed lines in Fig.\,\ref{Fig6}). 
Indeed there is no significant difference between the 
sequence of the segment sizes $\ell_n$ for different total 
lengths of the injected wire. This shows that the previously 
formed segments are not considerably affected during the 
compaction process, evidencing that the hierarchical 
folding events rarely happen.

In summary, we reported the compaction of 1D objects in 
spherical containers and showed how the morphology evolves 
from ordered (coiling) to disordered (folding and bending) 
structures in the phase space spanned by friction, torsion, 
and plasticity. The disorder reduces the compaction efficiency 
and causes a nontrivial system-size dependence, which is explained 
by self-avoiding random walks (SAWs) in confined geometries. 
Monitoring the evolution of segment-length distributions in 
highly disordered structures of plastic frictional wires showed 
that the compaction process is correlated: the longer the injected 
wire, the stronger the spatial exclusion effects leading to 
shorter segments. The self-avoidance constraint is only 
partially fulfilled due to sliding of the wires, leading to 
an exponent $\beta$ slightly larger than 5/3 as reported for 
SAWs. Our results provide new insight into underlying 
mechanisms of crumpling beyond the simple hierarchical 
description of the process, which also helps to better 
understand the reverse processes, e.g.\ viral DNA ejection 
\cite{Marenduzzo13} and unpacking of crumpled wires 
\cite{Sobral15}. While more detailed morphological 
information can be obtained via Discrete-Element Method 
(DEM) simulations, these are however computationally 
expensive. Our proposed sampling method opens the door 
to relatively simple Monte Carlo simulations of SAWs inside 
arbitrary confinements to obtain some of the macroscopic 
quantities of interest e.g.\ the arbitrary moments of 
segment-size distribution. 

\smallskip\smallskip

\vspace{0.8em}
{\footnotesize\noindent
{\sffamily\textbf{Methods.}}\vspace{1ex}

\noindent \textbf{Experimental setup.} The experimental setup 
consists of a rigid hollow spherical container of inner radius 
$R$ with a small hole to insert the wire (see Fig.\,\ref{Fig1}). 
Several transparent polymeric molds with radii $R{\in}[4,30\text{mm}]$ 
were used. A small nozzle and two counterrotating rollers were 
attached to the injecting hole to facilitate the control of the 
insertion speed.

\smallskip\smallskip

\noindent \textbf{Material properties.} As a model elastoplastic 
material, we chose solder wire $\text{Sn}_{60}\text{Pb}_{40}$ with 
Young's modulus $Y{\approx}30\,\text{GPa}$ and yield stress 
$\sigma{\approx}28\,\text{MPa}$. For the elastic wire experiments 
we mainly used fishing line with Young's modulus $Y{=}2.00{\pm}0.01
\,\text{GPa}$ (obtained experimentally by tensile tests). Moreover, 
elastic silicon wires and cotton threads with relatively lower Young's 
moduli $Y{\approx}5.0$ and $0.8\,\text{MPa}$ were also used.
The wire-wire and container-wire friction coefficients, using 
smooth wires and container walls, were $\mu_\text{ww}{=}0.20{\pm}0.02$ 
and $\mu_\text{cw}{=}0.40{\pm}0.02$, respectively. By roughening 
the plastic wires and the inner surface of the polymeric molds with 
sandpaper we obtained higher friction coefficients $\mu_\text{ww}{=}
0.45{\pm}0.02$ and $\mu_\text{cw}{=}0.45{\pm}0.02$. We also used 
smooth lubricated plastic wires and inner surfaces of the molds 
to lower the friction coefficients, leading to $\mu_\text{ww}
{=}0.12{\pm}0.02$ and $\mu_\text{cw}{=}0.18{\pm}0.02$. The lubrication 
was done with silicon oil. 

\smallskip\smallskip

\noindent \textbf{Insertion process and imaging.} We inserted wires of 
radius $r\,{=}\,0.4$, $0.5$, $0.6$, $0.8$, $\text{or}\,1.4\,\text{mm}$ 
into the molds with a slow feeding speed of about $1\,\text{mm s}^{-1}$ 
to avoid inertial effects. We checked that the results are independent 
of the feeding speed in the quasi-static compaction regime. The 
insertion process continued with the constant speed until 
the insertion force exceeded the power threshold of the motor 
and the wire buckled outside of the container. The final 
plastic-wire structure preserves its shape after opening 
the mold allowing for a detailed analysis of morphological 
changes, which can be considered as plastic deformations. 
In the low-torsion elastic setup, we allowed axial rotation of the 
wire between the nozzle and the sphere. The images 
presented in figures 1 and 2 were taken by a camera with 
pixel resolution $70 \mu\text{m}$. Before opening the 
molds, we filled them with a transparent gel in the case 
of elastic wires to preserve the shape of the final structure.

\smallskip\smallskip

\vspace{0.8em}
{\footnotesize\noindent
{\sffamily\textbf{Data availability.}}\vspace{1ex} The data that support 
the findings of this work are available from the corresponding authors 
on request.

\bigskip

\noindent\textbf{\sffamily Acknowledgments}

\noindent{This work was supported by DFG within SFB 1027 (A7). 
M.\,H.\, acknowledges support from NWO-VIDI grant 
680-47-548/983.}

\smallskip\smallskip

\noindent\textbf{\sffamily Author~contributions}

\noindent{M.\,R.\,S., M.\,H. and D.\,B. designed the research. A.\,F.,
J.\,N. and M.\,H. performed the experiments. M.\,R.\,S. 
developed the SAW model and performed simulations. All 
authors contributed to the analysis and interpretation 
of the results. M.\,R.\,S. drafted the manuscript.}

\smallskip\smallskip

\noindent\textbf{\sffamily Additional information}

\noindent{\bf Competing financial interests:} The authors declare no competing financial interests.

\end{document}